\documentclass[fleqn,usenatbib]{mnras}
    \usepackage{newtxtext,newtxmath}
    \usepackage[T1]{fontenc}
    \usepackage{booktabs}
    \usepackage{multirow} 
    \usepackage{makecell}
    \usepackage{graphicx}
    \usepackage{longtable}
    \usepackage{amsmath}
    \usepackage{lineno}

    \graphicspath{{./}}

    \newcommand{\CIAO}{{\fontfamily{qcr}\selectfont CIAO}}          
    \newcommand{\XSPEC}{{\fontfamily{qcr}\selectfont XSPEC}}        

    \newcommand{\pow}{{\fontfamily{qcr}\selectfont pow}}            
    \newcommand{\constant}{{\fontfamily{qcr}\selectfont constant}}  
    \newcommand{\agaus}{{\fontfamily{qcr}\selectfont agaus}}        

    \newcommand{\Ka}{{\fontfamily{qcr}\selectfont K$\alpha$}}   

    \newcommand{\Chandra}{{\fontfamily{qcr}\selectfont Chandra}}    
    \newcommand{\MEG}{{\fontfamily{qcr}\selectfont MEG}}            
    \newcommand{\HETG}{{\fontfamily{qcr}\selectfont HETG}}          
    \newcommand{\HEG}{{\fontfamily{qcr}\selectfont HEG}}            

    \newcommand{\ACISSHETG}{{\fontfamily{qcr}\selectfont ACIS-S HETG}}          

    \newcommand{\combinegratingspectra}{{\fontfamily{qcr}\selectfont combine$\_$grating$\_$spectra}}    
    \newcommand{\chandrarepro}{{\fontfamily{qcr}\selectfont chandra$\_$repro}}                          

    \newcommand{\logT}{$\log(T/\mathrm{K})$}            

    \newcommand{\Nion}{$N_\mathrm{ion}$}            %

    \newcommand{\chidof}{{\fontfamily{qcr}\selectfont $\chi^2/\mathrm{d.o.f}$}}     

    \newcommand{\NeX}{{\fontfamily{qcr}\selectfont Ne$_{\mathrm{X}}$}}                  
    \newcommand{\SXVI}{{\fontfamily{qcr}\selectfont S$_{\mathrm{XVI}}$}}                
    \newcommand{\SiXIV}{{\fontfamily{qcr}\selectfont Si$_{\mathrm{XIV}}$}}              

    \newcommand{\SiII}{{\fontfamily{qcr}\selectfont Si$_{\mathrm{II}}$}}              
    \newcommand{\SiIII}{{\fontfamily{qcr}\selectfont Si$_{\mathrm{III}}$}}              
    \newcommand{\SiIV}{{\fontfamily{qcr}\selectfont Si$_{\mathrm{IV}}$}}              
    \newcommand{\OVI}{{\fontfamily{qcr}\selectfont O$_{\mathrm{VI}}$}}              
    \newcommand{\OVII}{{\fontfamily{qcr}\selectfont O$_{\mathrm{VII}}$}}              
    \newcommand{\OVIII}{{\fontfamily{qcr}\selectfont O$_{\mathrm{VIII}}$}}              
    \newcommand{\NeIX}{{\fontfamily{qcr}\selectfont Ne$_{\mathrm{IX}}$}}              

    \title[Where is the Supervirial hot gas?]{Where is the Supervirial hot gas? I: A pilot study with sightlines to Galactic X-ray binaries}

    \author[Lara-DI et al.]{Armando Lara-DI$^{1},$\thanks{E-mail: alara@astro.unam.mx}
    Yair Krongold$^{1}$, 
    Smita Mathur$^{2,3}$, 
    Manami Roy$^{3}$,
    Rebecca L. McClain$^{2}$, 
    \newauthor
    Sanskriti Das$^{4}$, 
    Anjali Gupta$^{5}$
    \\
    $^{1}$Instituto de Astronomía, Universidad Nacional Autónoma de México, 04510 Mexico City, Mexico\\
    $^{2}$Department of Astronomy, The Ohio State University, 140 West 18th Avenue, Columbus, OH 43210, USA\\
    $^{3}$Center for Cosmology and Astro-Particle Physics, The Ohio State University, 191 West Woodruff Avenue, Columbus, OH 43210, USA\\
    $^{4}$Kavli Institute for Particle Astrophysics \& Cosmology, Stanford University, 452 Lomita Mall, Stanford, CA 94305, USA\\
    $^{5}$Columbus State Community College, 550 E Spring Street, Columbus, OH 43210, USA
   }

\date{Accepted XXX. Received YYY; in original form ZZZ}

\pubyear{2023}
\begin{document}
\label{firstpage}
\pagerange{\pageref{firstpage}--\pageref{lastpage}}
\maketitle

%
%

    \begin{abstract}

       Hot, \logT\ $\sim$ 7.5,  gas was recently discovered in the Milky Way in extragalactic sightlines. In order to determine its location, here we present sightlines to Galactic X-ray binaries (XRBs) passing through the Interstellar Medium (ISM). In this pilot study we investigate absorption features of \SXVI, \SiXIV, and \NeX\ in the spectra of three XRBs, namely 4U 1735-44, 4U 1820-30, and Cyg X-2, using Chandra High Energy Transmission Grating archival observations. We do not detect any of these lines. {We determine the 2$\sigma$ upper limit for the equivalent widths of the undetected absorption lines and the column densities of the corresponding ions.} We note that the 2$\sigma$ upper limits for \SXVI\ \Ka\ and \SiXIV\ \Ka\ are an order of magnitude smaller than those  previously detected in the extragalactic sightlines. Our finding suggests that if any gas at \logT\ $>7$ is present in the Galactic ISM, it is unlikely to be ubiquitous. This is an important result because it implies that \SXVI, \SiXIV\ {and \NeX\ }absorption detected in extragalactic sightlines is not from the ISM, but is likely from a hot gas phase in the extraplanar region beyond the ISM or in the extended CGM.

    \end{abstract}

%
%


%
%

    \section{Introduction \label{sec:intro}}

        Spiral galaxies are characterized by a prominent disk populated with stars. Between the stars resides the Interstellar Medium (ISM), a complex multi-phase environment comprising a mixture of dust, molecules, and atoms (e.g., \citealp{Rogantini2021} and references therein). 

        Studies {of} the Milky Way's ISM have shown that it comprises of a wide range of gas phases, from neutral atomic and molecular hydrogen to heavier ionized elements such as \SiII, \SiIII, \SiIV, \NeIX, \OVI, and \OVII\ (e.g., \citealp{Werk2019} and references therein; \citealp{Rogantini2021}). Energetic phenomena like stellar winds, supernova explosions, and interactions between binary star systems can both ionize and disperse the elements in the ISM.     

        Above and below the galactic disk, we find a much more diffuse and ionized medium called the Circumgalactic Medium (CGM) (see \citealp{Mathur2022} for a review). The CGM is defined as the gas extending beyond the galactic disk and inside the Galactic virial radius. Elements in different ionization states, such as \OVI, \OVII, \OVIII, and \NeIX\ have been found in the CGM of the Milky Way (e.g., \citealp{Gupta2012}; \citealp{Mathur2012}; \citealp{Tumlinson2017}; \citealp{Das2019}; \citealp{Bhattacharyya2022}; \citealp{LaraDI2023}). 
        
        Unlike the Galactic disk's energetic processes explaining the observed ionized elements in the ISM, the CGM lacks a precise, direct mechanism for producing and ionizing the gas. Theoretical simulations show that the CGM is shock heated during the collapse of material in galaxy formation (see \citealp{Crain2023} and references therein). Numerical simulations have proposed that feedback processes from starbursts and active galactic nuclei expelled ionized metals from the disk into the CGM (e.g., \citealp{Stinson2012}), {playing a significant role in shaping the CGM}.

        The Milky Way's CGM is composed of two or three dominant phases: one at virial temperature (\logT\ $\sim$ 6) \citealp{Gupta2012} and references therein, a hotter component at super-virial temperature (\logT\ $\sim$ 7 - 7.5; e.g., \citealp{Das2019}; \citealp{Das2021}; \citealp{Gupta2022}; \citealp{Bhattacharyya2022}), and a sub-virial component {at \logT $\sim$5.5 (\citealp{Das2021})}. The \logT\ $\sim$ 6 component has been known for over {two decades}, but the \logT\ $\sim$ 7 has been recently discovered in the last few years (\citealp{Das2019}).
        
        In a recent study using stacked X-ray spectral observations towards tens of different extragalactic sightlines, \citet{LaraDI2023} detected \SiXIV\ \Ka\ and \SXVI\ \Ka\ in the CGM of the Milky Way. These lines are associated {with} a gas phase with temperature \logT\ $\sim$ 7.5. Gas at these temperatures is not predicted by the simulations of galaxy formation (e.g., \citealp{Stinson2012}), and its detection raises fundamental questions: Does this gas component arise from the CGM or the ISM of our Galaxy? What is the role of the Galactic CGM in the evolution of the Galaxy? {What could be the origin of such supervirial gas, and how is this gas interacting with the Galaxy?}
        
        Building on our understanding of ionized gas in the Milky Way, in the present work we analyze Chandra's High-Energy Transmission Grating (\HETG) X-ray observations toward three Galactic X-ray Binaries (XRBs): 4U 1735-44, 4U 1820-30, and Cyg X-2. This is a pilot study; in future works we will present spectra from a large sample of XRBs. Our goal is to investigate \NeX, \SiXIV, and \SXVI\ absorption that has been previously detected in the extragalactic sightlines (e.g., \citealp{Das2019}; \citealp{Das2021}; \citealp{LaraDI2023}; \citealp{McClain2024}; {\citealp{LaraDI2024a}.; Roy et al. 2024 in prep.}) {and determine whether the absorption occurs in the ISM of the Galaxy, or {it is} truly from extended regions beyond the ISM}. 

        We structure this paper as follows. In Section 2, we present the data sample. In Section 3, we describe the data analysis. In Section 4, we present our results. In Section 5, we discuss their implications, and in Section 6 we present our conclusions.

%
%

    \section{Data Sample}
    
        We downloaded 23 (623.56 ks) archival \Chandra\ \ACISSHETG\ observations towards XRBs 4U 1735-44, 4U 1820-30, and Cyg X-2 (see Table~\ref{tab:listObsIDs}). {These particular sources are 3 of the 41 Galactic XRBs with high signal-to-noise ratio (S/N) ($>$ 50) in the Chandra HETG database. A larger XRB sample will be used in a future  paper (in preparation).} We reprocessed these observations using \chandrarepro\ script with the \Chandra\ Interactive Analysis of Observations\footnote{https://asc.harvard.edu/ciao/} (\CIAO) software (v4.13).

        \begin{table*}
            \centering
            \caption{ObsID List}
            \label{tab:listObsIDs}
                \resizebox{\textwidth}{!}{%
            \begin{tabular}{ccccccc}
            \hline
           Target & Galactic Lon [deg] & Galactic Lat [deg] & Distance [kpc] & Exp. Mode & Exp. Time (ks) & ObsIDs \\
           (1) & (2) & (3) & (4) & (5) & (6) \\
            \hline
            \multirow{2}{*}{4U 1735-44} & \multirow{2}{*}{346.053614} & \multirow{2}{*}{-6.993435} & \multirow{2}{*}{9.2$^a$} & CC & 47.12 & 6637, 6638 \\
                                        & & & & TE & 24.35 & 704 \\
            \hline
            \multirow{3}{*}{4U 1820-30} & \multirow{3}{*}{2.788145} & \multirow{3}{*}{-7.913689} & \multirow{3}{*}{6.4$^b$}  &  \multirow{2}{*}{CC}  & \multirow{2}{*}{299.15}   & 663, 6634, 7032, 22276, 22277, 24698, 25029    \\
                                        & & & &                       &                           & 25030, 25031, 25032, 25033 ,25037, 25038       \\
                                        & & & &  \multirow{1}{*}{TE}  & \multirow{1}{*}{20.29}    & 1021, 1022                                     \\
            \hline
            \multirow{2}{*}{Cyg X-2}    & \multirow{2}{*}{87.32845} & \multirow{2}{*}{-11.316192} & \multirow{2}{*}{7.2$^c$}  &  CC                    & 189.38                    & 8170, 8599, 10881                             \\
                                        & & & &  TE                    & 43.27                     & 1102, 1016                                    \\
            \hline
           TOTAL              & & & &                        & 623.56                  &\\
           \hline
            \multicolumn{7}{l}{$^a$ \citealp{VanParadijs1995}}\\
            \multicolumn{7}{l}{$^b$ \citealp{Vacca1986}}\\
            \multicolumn{7}{l}{$^c$ \citealp{Orosz1999}}\\
            \end{tabular}}            
        \end{table*}

        We stacked the observations using the \combinegratingspectra\ command in the \CIAO\ software. The stacking was performed in order to get four stacked spectra for each XRB: one for the High Energy Gratings (\HEG) continuous clocking readout mode (CC) observations, the second one for the \HEG\ timed exposure readout mode (TE) observations, the third for the Medium Energy (\MEG) CC observations, and the last one for the {\MEG}\ TE observations. {The resolving power ($\lambda/\Delta \lambda$) for \HEG\ is 1000 at 12.4 \AA, while in \MEG\ is 540 at this same wavelength.} In Table~\ref{tab:SN} we show the S/N per resolution element (SNRE) for CC (column 3) and TE (column 4) at 4.729 \AA. At this wavelength, corresponding to the rest-wavelength of \SXVI\ \Ka, the SNRE is $>$ 20. {We note that the CC mode observations have larger exposure times, resulting in better SNRE, compared to TE mode observations.} 

        {We note that CC mode observations in all three XRBs have a higher exposure time than TE. Additionally, in 5 out of the 6 MEG/HEG stacked observations, CC has a higher count rate than TE. Therefore, the signal-to-noise ratio is higher in CC mode observations than in TE.}

        In the spectral range 4 - 19 \AA, TE observations can be affected by pileup when the sources are observed in a high flux state (e.g., \citealp{Rogantini2021}). {Pileup occurs when multiple photons are detected as a single event, leading to a loss of information and energy spectral distortion. Narrow features, corresponding to specific energy lines, are less affected by pileup due to the lower likelihood of multiple photons coincidentally arriving at the exact energy levels needed for these features. Consequently, narrow features are less influenced by pileup compared to broader features in the observed X-ray spectra,} as also noted by \citet{Rogantini2021}.  Therefore, we analyze both the TE and CC spectra of all the three targets.

        {The closest Galactic sightline in this work to an extragalactic sightline where super-virial hot gas has been detected in absorption has an angular separation of 54.7 degrees. This corresponds to the angular distance between 4U 1820-30 and 1ES 1553+113 (\citealp{Das2019}). Given this large separation, we cannot directly comment on the location of the hot gas in any particular extragalactic sightline. }

 \begin{table}
            \centering
            \caption{Signal to Noise ratio (S/N) {per resolution element (SNRE {; 23 m\AA\ for \MEG, and 12 m\AA\ for \HEG}) at 4.729 \AA} in CC (column 3) and TE (column 4) {spectra}.}
            \label{tab:SN}
                \resizebox{0.3\textwidth}{!}{%
            \begin{tabular}{cccc}
            \hline
            Target & Grating & CC & TE \\
            (1) & (2) & (3) & (4) \\
            \hline
            \multirow{2}{*}{4U 1735-44} & \HEG & 29 & 21 \\
                                        & \MEG & 61 & 40 \\
            \hline
            \multirow{2}{*}{4U 1820-30} & \HEG & 75 & 19 \\
                                        & \MEG & 162 & 43 \\
            \hline
            \multirow{2}{*}{Cyg X-2}    & \HEG & 115 & 33 \\
                                        & \MEG & 212 & 72 \\
            \end{tabular}
            }
        \end{table}
        
%
%

    \section{Analysis \label{sec:analysis}}

        We used the software \XSPEC\footnote{https://heasarc.gsfc.nasa.gov/docs/software/heasoft/} (v12.13.0) and $\chi^{2}$ statistics to perform the spectral fitting of the CC \MEG\ and \HEG\ observations of the three XRBs, analyzing six spectra simultaneously. We did the same for the TE observations.
            
        We constrained our modeling to the spectral region around $\pm$ 0.25 \AA\ from the expected position of the rest-wavelength of each ionic transition of interest, i.e., \SXVI\ \Ka\ ($\lambda$ 4.729 \AA), \SiXIV\ \Ka\ ($\lambda$ 6.182 \AA), and \NeX\ \Ka\ ($\lambda$ 12.134 \AA). 

        To begin with, we fit the local continuum in these small ranges using a power-law (\pow). The \pow\ parameters between \MEG\ and \HEG\ of the same observation were set to be the same; however, if necessary, we used the \constant\ parameter to account for any  difference in the counts between the data. On the other hand, the \pow\ of the three XRBs was allowed to be different. Additionally, if the spectra presented another feature to the one we were looking for, we added a Gaussian profile (\agaus) to account for it.
        
        Then, with a narrow Gaussian profile at the rest wavelength of \SXVI\ \Ka, \SiXIV\ \Ka, and \NeX\ \Ka\, we modeled the possible presence of an absorption feature. We set the position of each Gaussian profile to the expected value of the line and let it vary  $\pm$ 0.012 \AA, corresponding to the resolution element of \MEG\ and two times the resolution element of \HEG. Finally, we fixed the line width to zero \AA\ {(since the lines are unresolved with the \HETG\ resolution)} and let the normalization vary. The line's position and normalization were forced to be the same in all the six spectra of each XRB.

%
%

    \section{Results} \label{sec:results}    

    In the line of sight towards 4U 1735-44, 4U 1820-30, and Cyg X-2, we do not detect any absorption line at $z\approx0$ corresponding to \NeX, \SiXIV, or \SXVI.  Since the fit statistically does not require any of these lines, we have determined the 2$\sigma$ upper limit for the equivalent width (EW) and ionic column density (\Nion) of any undetected ISM absorption. Our results are shown in Table~\ref{tab:results}, which includes (1) the exposure mode, (2) the expected ionic transition, (3) the center of the undetected absorption line, (4) the 2$\sigma$ upper limit of the EW (m\AA), (5) the 2$\sigma$ upper limit of the \Nion\ calculated using the curve of growth, (6) the \chidof\ statistics of the fit modeling the linear part of the local continuum, (7) and the \chidof\ statistics of the fit when an absorption feature is considered in the spectra at the expected ionic transition (hereafter c+l).
    
    In each absorption feature, the 2$\sigma$ upper limit of the EW is below 0.5 m\AA. We note  that these values are consistently greater in TE than in CC data, consistent with the better S/N in the CC spectra.
    
    We also performed the analysis separately on  the individual spectra of each X-ray binary. We did not {detect} any lines, consistently {with} the results reported above. {In Table~\ref{tab:resultsall}, we show the 2$\sigma$ upper limit of the EW (m\AA) and \Nion\ for each sightline when we analyzed each spectrum separately.} {In Figure~\ref{fig:plot_p3} we show MEG (CC) spectra around \NeX\ \Ka, \SiXIV\ \Ka, and \SXVI\ \Ka\ toward each sightline, with the line upper limit superposed (red line). For comparison, we overplotted absorption lines {stronger by a factor of 3} 
     (dashed green line). {These stronger features are not lost in the noise in most of the panels in Figure~\ref{fig:plot_p3}.} 
      {In MEG, the resolution element is $\Delta\lambda$ = 0.023 \AA, while the pixel scale is 0.01112 \AA\ per pixel, resulting in approximately 2.06 pixels per resolution element. However, each channel in the MEG spectrum is separated by $\Delta\lambda$ = 0.005 \AA, leading to about 4.6 MEG channels per resolution element. Therefore, for visual purposes only, we have binned the spectra in Figure~\ref{fig:plot_p3} by a factor of 5, corresponding to approximately 2 pixels. This is also consistent with the SNRE reported in Table~\ref{tab:SN}.}

    \begin{table*}
        \centering
        \caption{Results. Here we present the 2$\sigma$ upper limits of absorption features observed in the positions of the ionic transitions. Spectra fitting was performed simultaneously for three XRB sightlines instead of stacking them. {Results listed for CC are for MEG (CC) data shown in Fig.~\ref{fig:plot_p3}}}
        \resizebox{\textwidth}{!}{%
        \begin{tabular}{lllllll}
        Exposure Mode & Ionic Transition & Wavelength (\AA) & EW(m\AA) & \Nion\ ($\mathrm{cm}^{-2}$)   &  $\chi^2$/d.o.f & $\chi^2_{c+l}$/d.o.f$_{c+l}$ \\
        (1) & (2) & (3) & (4) & (5) & (6)  & (7)  \\
        \hline
     \multirow{3}{*}{CC} & \NeX\    ($\lambda$12.134)   & 12.122    &  $<0.15$ & $<3.18\times10^{14}$  & 903.48/894  & 899.71/892   \\
                         & \SiXIV\  ($\lambda$6.182)    & 6.176     &  $<0.05$  & $<4.09\times10^{14}$  & 1011.10/894 & 1010.84/892  \\
                         & \SXVI\   ($\lambda$4.729)    & 4.717     &  $<0.07$  & $<9.78\times10^{14}$ & 1071.52/892 & 1065.64/894  \\ \hline
     \multirow{3}{*}{TE} & \NeX\    ($\lambda$12.134)   & 12.125    &  $<0.38$ & $<8.07\times10^{14}$  & 813.88/894  & 809.22/892   \\
                         & \SiXIV\  ($\lambda$6.182)    & 6.185     &  $<0.14$  & $<1.14\times10^{15}$  & 1025.32/894 & 1020.49/892  \\
                         & \SXVI\   ($\lambda$4.729)    & 4.726     &  $<0.21$ & $<2.93\times10^{15}$  & 887.11/894  & 886.11/892   \\
    \end{tabular}
    }
    \label{tab:results}
    \end{table*}


\begin{table*}\label{tab:resultsall}
\centering
\caption{2$\sigma$ upper limit for the EW and \Nion\ for each sightline.}
    \begin{tabular}{ccclcc}
Target	&	Mode	&	Instrument	&	Line	&	2$\sigma$ EW [m\AA]	&	2$\sigma$ \Nion\ [cm$^{-2}$]	\\ 
(1) & (2) & (3) & (4) & (5) & (6) \\ \hline \hline
4U 1735-44	&	CC	&	MEG	&	\SiXIV	&	$<$0.54	&	$<$4.42e+15	\\
4U 1735-44	&	CC	&	MEG	&	\SXVI	&	$<$0.78	&	$<$1.09e+16	\\
4U 1735-44	&	CC	&	MEG	&	\NeX	&	$<$1.72	&	$<$3.65e+15	\\
4U 1735-44	&	CC	&	HEG	&	\SXVI	&	$<$0.86	&	$<$1.20e+16	\\
4U 1735-44	&	CC	&	HEG	&	\SiXIV	&	$<$0.55	&	$<$4.50e+15	\\
4U 1735-44	&	CC	&	HEG	&	\NeX	&	$<$2.49	&	$<$5.28e+15	\\
4U 1735-44	&	TE	&	HEG	&	\SXVI	&	$<$1.29	&	$<$1.80e+16	\\
4U 1735-44	&	TE	&	HEG	&	\SiXIV	&	$<$0.83	&	$<$6.79e+15	\\
4U 1735-44	&	TE	&	HEG	&	\NeX	&	$<$2.89	&	$<$6.13e+15	\\
4U 1735-44	&	TE	&	MEG	&	\SXVI	&	$<$1.19	&	$<$1.66e+16	\\
4U 1735-44	&	TE	&	MEG	&	\SiXIV	&	$<$0.88	&	$<$7.20e+15	\\
4U 1735-44	&	TE	&	MEG	&	\NeX	&	$<$2.05	&	$<$4.35e+15	\\ \hline
4U 1820-30	&	CC	&	MEG	&	\SXVI	&	$<$0.29	&	$<$4.05e+15	\\
4U 1820-30	&	CC	&	MEG	&	\SiXIV	&	$<$0.19	&	$<$1.55e+15	\\
4U 1820-30	&	CC	&	MEG	&	\NeX	&	$<$0.64	&	$<$1.36e+15	\\
4U 1820-30	&	CC	&	HEG	&	\SXVI	&	$<$0.28	&	$<$3.91e+15	\\
4U 1820-30	&	CC	&	HEG	&	\SiXIV	&	$<$0.22	&	$<$1.80e+15	\\
4U 1820-30	&	CC	&	HEG	&	\NeX	&	$<$0.81	&	$<$1.72e+15	\\
4U 1820-30	&	TE	&	MEG	&	\SXVI	&	$<$1.12	&	$<$1.57e+16	\\
4U 1820-30	&	TE	&	MEG	&	\SiXIV	&	$<$0.81	&	$<$6.62e+15	\\
4U 1820-30	&	TE	&	MEG	&	\NeX	&	$<$2.04	&	$<$4.33e+15	\\
4U 1820-30	&	TE	&	HEG	&	\SXVI	&	$<$1.13	&	$<$1.58e+16	\\
4U 1820-30	&	TE	&	HEG	&	\SiXIV	&	$<$0.81	&	$<$6.62e+15	\\
4U 1820-30	&	TE	&	HEG	&	\NeX	&	$<$2.14	&	$<$4.54e+15	\\ \hline
Cyg X-2	&	CC	&	MEG	&	\SXVI	&	$<$0.22	&	$<$3.07e+15	\\
Cyg X-2	&	CC	&	MEG	&	\SiXIV	&	$<$0.14	&	$<$1.14e+15	\\
Cyg X-2	&	CC	&	MEG	&	\NeX	&	$<$0.38	&	$<$8.07e+14	\\
Cyg X-2	&	CC	&	HEG	&	\SXVI	&	$<$0.18	&	$<$2.52e+15	\\
Cyg X-2	&	CC	&	HEG	&	\SiXIV	&	$<$0.16	&	$<$1.31e+15	\\
Cyg X-2	&	CC	&	HEG	&	\NeX	&	$<$0.60	&	$<$1.27e+15	\\
Cyg X-2	&	TE	&	MEG	&	\SXVI	&	$<$0.64	&	$<$8.94e+15	\\
Cyg X-2	&	TE	&	MEG	&	\SiXIV	&	$<$0.53	&	$<$4.33e+15	\\
Cyg X-2	&	TE	&	MEG	&	\NeX	&	$<$1.07	&	$<$2.27e+15	\\
Cyg X-2	&	TE	&	HEG	&	\SXVI	&	$<$0.61	&	$<$8.52e+15	\\
Cyg X-2	&	TE	&	HEG	&	\SiXIV	&	$<$0.47	&	$<$3.84e+15	\\
Cyg X-2	&	TE	&	HEG	&	\NeX	&	$<$1.53	&	$<$3.25e+15	\\
    \end{tabular}
\end{table*}

    \begin{figure*}
        \centering
        \renewcommand{\arraystretch}{2}\addtolength{\tabcolsep}{1pt}
        \includegraphics{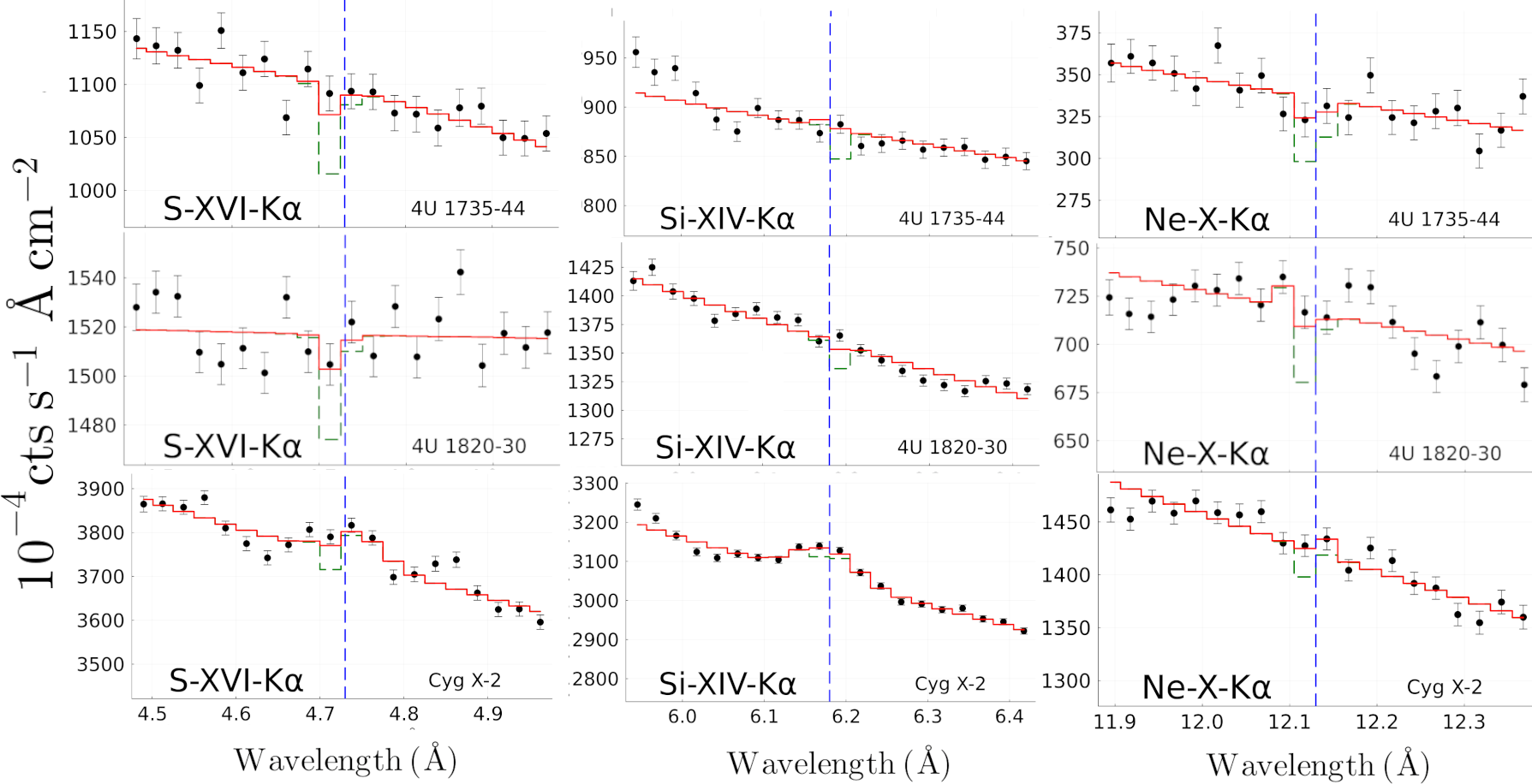}
        \caption{{Here we display the model including the $2\sigma$ upper limit of the normalization of absorption lines (red curve, Table~\ref{tab:resultsall}) best-fitted to the MEG-CC data (black). We show for each sightline the position of \NeX\ \Ka, \SiXIV\ \Ka, and \SXVI\ \Ka\ at $z=0$ (vertical blue dashed line). In dashed {green} are the absorption features superposed if they were three times stronger. {Data points shown are binned by a factor of 5, and show resolution elements, not pixels or channels.}}}
        \label{fig:plot_p3}
    \end{figure*}

%
%

\section{Discussion \label{sec:discussion}}
        
        Different studies have characterized a hot gas phase at \logT\ $\sim$ 6 in the ISM of the Milky Way (e.g., \citealp{Rogantini2021}). Consistently, \NeIX, arising from this gas phase, has been detected towards the sightlines analyzed here (e.g., \citealp{Pinto2013} for 4U 1735-44; \citealp{Cackett2008} for 4U 1820-30; \citealp{Yao2009} for Cyg X-2). On the other hand, \NeX, \SiXIV, and \SXVI\ are charge states tracing hotter gas. We do not detect these ions in the spectra of these XRBs. Furthermore, we are not aware of any detection of these charge states in the Galactic ISM in literature.  This suggests that while gas at \logT\ $\sim$ 6 is commonly found in the ISM, hotter gas is not. 

        The 2$\sigma$ upper limits for the EW and \Nion\ of the undetected ISM absorption features at the rest-wavelength of \SXVI\ \Ka\ and \SiXIV\ \Ka\ are at least an order of magnitude smaller than those measured in the extragalactic sightlines (e.g., \citealp{Das2021}; \citealp{LaraDI2023}). 
        Similarly, the 2$\sigma$ upper limits for \Nion\ of \NeX\ are an order of magnitude smaller than the detection column densities in \cite{Das2019, Das2021} and \cite{McClain2024}. 
        Therefore, the ISM contribution to the super-virial hot component detected in the extragalactic sightlines is negligible. 

        The above result is derived from only three XRB sightlines in this pilot study. In future works we will extend this to a large sample of XRBs to arrive at a firm conclusion. If the super-virial hot component is not in the ISM, as suggested here, it will have to be in regions beyond the ISM. It may be in the extra-planar region (e.g., \citealp{Vijayan2022}) or in the extended diffuse CGM . At this stage we cannot dientangle these two possibilities. 

        {In order to determine whether we can detect the weak ISM lines from the hot gas {(\logT$\sim$7 - 7.5)}, {if any}, with future observations, we simulated the XRISM spectra. We used the best fit model of the \SiXIV\ line from the \Chandra\ \HEG\ observation of Cyg X-2 in the CC mode. We assumed that the line strength is equal to the $3\sigma$ upper limit observed with HEG ($0.24$ m\AA). After fitting the same model to the simulated  XRISM data, we {can detect the \SiXIV\ \Ka\ line with $3\sigma$ significance} with a $60$ ks exposure time. This is $< 1/3$ exposure time of \Chandra. If the actual line has smaller EW, the exposure time would be larger.

%
%

    \section{Conclusion} \label{sec:conclusions} 

        Our findings suggest that gas with super-virial temperatures is not commonly found in the Galactic ISM, and its contribution to the $z=0$ absorption detected in extragalactic sightlines is marginal, if any. The super-virial hot gas detected in the extragalactic sightlines is therefore likely to be present in the extraplanar region or in the extended CGM. In future works we plan on further disentangling the location of the hot gas. 

%
%

   \section{Acknowledgments} \label{sec:acknowledgments} 

        The authors of this work thank the anonymous referee for his/her valuable comments. L.D.I. acknowledges support from CONACYT through the PhD scholarship grant 760672. Y.K. acknowledges support from grant DGAPA PAPIIT 102023. S.M. is grateful for the grant provided by the National Aeronautics and Space Administration through Chandra Award Number AR0-21016X issued by the Chandra X-ray Center, which is operated by the Smithsonian Astrophysical Observatory for and on behalf of the National Aeronautics Space Administration under contract NAS8-03060. S.M. is also grateful for the NASA ADAP grant 80NSSC22K1121. S.D. acknowledges support from the KIPAC Fellowship of Kavli Institute for Particle Astrophysics and Cosmology Stanford University. 
        
    \section*{Data Availability}

        All data used in this article are available online via the Chandra Data Archive: https://cda.harvard.edu/chaser/
        
%
%

\bibliographystyle{mnras}
\bibliography{bibliography} 

%
%

\bsp	
\label{lastpage}
\end{document}